\documentclass[twocolumn,showpacs,superscriptaddress,preprintnumbers,amsmath,amssymb,epsfig,floatfix,prb]{revtex4}
\usepackage{amssymb}
\usepackage{epsfig}
\usepackage{calrsfs}
\DeclareMathAlphabet{\pazocal}{OMS}{zplm}{m}{n}
\usepackage{graphicx}
\usepackage{dcolumn}
\usepackage{color}
\usepackage{bm}
\usepackage[normalem]{ulem}
\usepackage{soul,xcolor}

\begin{document}
\setstcolor{red}
\title{Electronic structure and spin-orbit driven novel magnetism in $d^{4.5}$ insulator Ba$_3$YIr$_2$O$_9$}
\author{S. K. Panda }

\altaffiliation[]{Presently at Uppsala University, Uppsala, Sweden}
\affiliation{Centre for Advanced Materials, Indian Association for the Cultivation of Science, Jadavpur, Kolkata-700032, India}
\author{S. Bhowal}

\affiliation{Department of Solid State Physics, Indian Association for the Cultivation of Science, Jadavpur, Kolkata-700032, India}

\author{Ying Li}

\affiliation{Institut f\"ur Theoretische Physik, Goethe-Universit\"at Frankfurt, Max-von-Laue-Strasse 1, 60438 Frankfurt am Main, Germany}

\author{S. Ganguly}
\affiliation{Department of Solid State Physics, Indian Association for the Cultivation of Science, Jadavpur, Kolkata-700032, India}

\author{Roser Valent\'i}

\affiliation{Institut f\"ur Theoretische Physik, Goethe-Universit\"at Frankfurt, Max-von-Laue-Strasse 1, 60438 Frankfurt am Main, Germany}

\author{L. Nordstr\"om}

\affiliation{Department of Physics and Astronomy, Uppsala University, Box 516, SE-75120 Uppsala, Sweden}

\author{I. Dasgupta}

\altaffiliation{sspid@iacs.res.in}
\affiliation{Centre for Advanced Materials, Indian Association for the Cultivation of Science, Jadavpur, Kolkata-700032, India}

\affiliation{Department of Solid State Physics, Indian Association for the Cultivation of Science, Jadavpur, Kolkata-700032, India}


\begin{abstract}
We have carried out a detailed first-principles study of a d$^{4.5}$ quaternary iridate Ba$_3$YIr$_2$O$_9$ both in its 6H-perovskite-type ambient pressure (AP) phase and also for the high pressure (HP) cubic phase. Our analysis reveals that the AP phase belongs to the intermediate spin-orbit coupling (SOC) regime. This is further supported by the identification of the spin moment as the primary order parameter (POP) obtained from a magnetic multipolar analysis. The large $t_{2g}$ band width renormalizes the strength of SOC and the Ir intersite
exchange interaction dominates resulting in long range magnetic order in the AP phase. In addition to SOC and Hubbard $U$, strong intradimer coupling is found to be crucial for the realization of the insulating state. At high pressure (HP) the system undergoes a structural transformation to the disordered cubic phase. 
In sharp contrast to the AP phase, the calculated exchange interactions in the HP phase are found to be much weaker and SOC dominates leading to a quantum spin orbital liquid (SOL) state.\\
\end{abstract}

\maketitle
\par

The 5-$d$ Ir based oxide systems have become very fascinating as they offer a novel ground
for understanding the physics driven by the interplay between SOC,
on-site Coulomb repulsion ($U$), crystal-field
splitting, and intersite hopping. Many exotic phases like novel $j_{eff}$=$\frac{1}{2}$
Mott state, quantum spin liquid (QSL), topological insulator, and other emergent states have been realized in
these systems~\cite{Sr2IrO4_Insulator,Ba2IrO4_Insulator,Na2IrO3,QSL_Na4Ir3O8,QSL_Ba3IrTi2O9,TI_Na2IrO3_1,TI_Na2IrO3_2,IrO2}.
Kim \textit{et al.}~\cite{Kim1, Kim2} first reported spin-orbit driven Mott-insulating state in Sr$_2$IrO$_4$ where a large SOC ($\lambda$) further 
splits the $t_{2g}$ orbitals of Ir$^{4+}$ (5$d^5$) into a completely filled $j_{eff}$ = $\frac{3}{2}$ and a half-filled $j_{eff}$ = $\frac{1}{2}$
doublet which in the presence of relatively small value of $U$ split into lower and upper Hubbard bands leading to a spin-orbit-driven Mott insulator. In this respect, the 6H perovskite type quaternary iridates Ba$_3$MIr$_2$O$_9$, where Ir ions form structural dimers, attracted considerable attention
 because a non magnetic M atom provides a knob to tailor the valence of Ir. 
 Interestingly, Ir in oxidation state other than the usual 4+ (d$^5$) are not only insulating but also exhibit emergent phases like a spin-orbital liquid (SOL) state in Ba$_3$ZnIr$_2$O$_9$~\cite{ZnIr2} and a pressure induced transition from long range order to spin liquid state in Ba$_3$YIr$_2$O$_9$~\cite{Spinliquid_cubicphase}. The compound Ba$_3$YIr$_2$O$_9$ with fractional charge state Ir$^{4.5+}$ is intriguing on several counts. In the $j_{eff}$=$\frac{1}{2}$ picture the insulating state of Ba$_3$YIr$_2$O$_9$ is not obvious as Ir with 4.5 electrons is a quarter filled system as opposed to half filled $d^{5}$ iridates. The origin of pressure induced magnetic transition in this system has also remained unexplored. 
Further there could be a possibility of a temperature dependent structural transformation to a 
charge ordered (CO) state, as seen in Ba$_3$NaRu$_2$O$_9$~\cite{RuCharge_ordering}. 
\par
Earlier experimental studies on Ba$_3$YIr$_2$O$_9$~\cite{Doi} at ambient pressure (AP) reported a sharp anomaly at
4 K in the heat capacity data although no such feature was seen in the susceptibility data.
Later, Dey \textit{et al.}~\cite{TDey} found that the variation of the susceptibility with 
temperature shows a weak anomaly at about 4 K  in agreement
with previous heat capacity measurements reflecting the presence of long-range 
ordering. The susceptibility data~\cite{TDey} follows a Curie-Weiss (CW) behavior with $\theta_{CW}\sim0$ and effective moment 0.3 $\mu_B$ much less than the spin-only value expected for a $d^{4.5}$ system. On the other hand, the high-pressure (HP) synthesized sample undergoes a structural phase transition and does not order down to 2 K as evidenced from susceptibility, heat capacity, 
and nuclear magnetic resonance (NMR) measurements and is suggested to be a 5d-based, gapless QSL~\cite{Spinliquid_cubicphase}.
\par 
In view of the above, and also due to the fact that iridates, where Ir ions are in a fractional charge state and form structural dimers, have not been explored before, we have performed density functional theory (DFT) electronic structure calculations within the local density approximation (LDA) as well as LDA+U+SOC that includes the Hubbard $U$ and SOC using the augmented plane waves plus local orbitals (APW + lo) method as implemented in the ELK code~\cite{Elk1,Elk2,Elk3}. The SOC was treated through the second variational method. In this approach, given a value of $U$ using a screened Coulomb potential (Yukawa potential) the corresponding 
onsite exchange energy $J$ (Hund's coupling) is calculated~\cite{Elk2}.
 The ELK code was also employed to identify the primary order parameter (POP) responsible for breaking of the  time reversal (TR) symmetry. Calculations were also done within the generalized gradient approximation (GGA) ~\cite{gga} as well as GGA+U+SOC using the LAPW method as implemented in Wien2k~\cite{wien2k} and also projector augmented wave (PAW) method as implemented in the Vienna \textit{ab initio} simulation package (VASP)~\cite{PAW1, PAW2, vasp1, vasp2}. While these results are very similar to the LDA+U+SOC study, as expected, GGA systematically overestimated the magnitude of the magnetic moment. 
In order to extract the tight binding parameters and to understand the crystal field splitting, we have constructed low energy model Hamiltonian using the downfolding procedure as implemented in the N$^{th}$ order muffin-tin-orbital (NMTO) method~\cite{NMTO1, NMTO2, NMTO3} with self-consistent potentials obtained from linear muffin tin orbital (LMTO)~\cite{} calculations. The technical details of all the methods are discussed in Supplementary Materials (SM)~\cite{SM}. 
\par  

\begin{figure}[t]
\centering
\includegraphics[scale =.30]{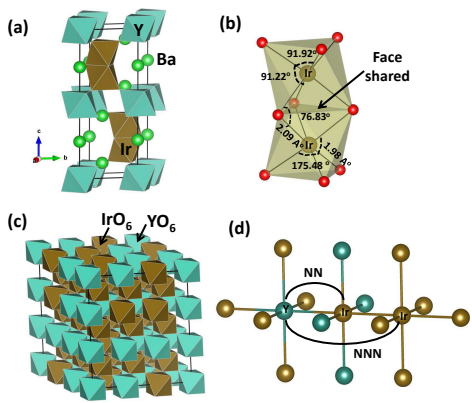}
\caption{(color online) (a) The crystal structure for the hexagonal AP phase of Ba$_3$YIrO$_9$. (b) Ir$_2$O$_9$ face-shared bioctahedra. (c) Crystal structure for the Cubic HP phase Ba$_{2}$Ir(Y$_{2/3}$Ir$_{1/3}$)O$_{6}$. (d) In the HP phase, the nearest neighbor(NN) of Y is always Ir while the next nearest neighbor(NNN) of Y can be either  Ir or Y due to disorder.} 
\label{fig1}
\end{figure}

At ambient pressure (AP), Ba$_3$YIr$_2$O$_9$ crystallizes in a 6H-perovskite-type structure~\cite{Spinliquid_cubicphase} with space group P6$_3$/mmc 
containing two formula units and hence two structural Ir dimers in the unit cell which are connected via O-Y-O paths along the c-axis [see Fig~\ref{fig1}(a)].
As shown in Fig~\ref{fig1}(b), each distorted IrO$_6$ octahedra share their face
with the neighboring one, forming Ir$_2$O$_9$ face-shared bioctahedra. The high pressure (HP) phase of Ba$_3$YIr$_2$O$_9$, on the other hand crystallizes in the disordered cubic phase Ba$_{2}$Ir(Y$_{2/3}$Ir$_{1/3}$)O$_{6}$ as shown in Fig.~\ref{fig1}(c),
where IrO$_6$ (brown) and YO$_6$ (cyan) octahedra are arranged, in such a way that each Y has 6 Ir as its NN but the NNN can be either Y or Ir [see Fig.~\ref{fig1}(d)]~\cite{Spinliquid_cubicphase}.
\par
 We have first investigated the electronic structure of the AP phase of Ba$_3$YIr$_2$O$_9$ without any magnetic order. The characteristic feature of the band structure (see Fig. 1 of SM) is the isolated manifold of twelve $t_{2g}$ bands hosting the Fermi level (E$_F$) arising from the four Ir atoms in the unit cell and is well separated by a gap of 2.5 eV from another complex of eight $e_g$ bands. The $t_{2g}$-$e_{g}$ crystal field splitting ($\Delta_{t_{2g}-e_{g}}$) is large. The distortion of the IrO$_6$ octahedron  further splits the $t_{2g}$ states into singly degenerate $a_1$ and doubly degenerate $e_1$ states with the $e_1$ states lower in energy by $\Delta_{CF}$ = 0.16 eV (see SM for details).
The Ba-$s$ and Y-$d$ states are completely empty and hence lie above E$_F$ while O-$p$ states, admixed with Ir-$d$ states, are mostly occupied consistent with the nominal ionic formula Ba$_3^{2+}$Y$^{3+}$Ir$_2^{4.5+}$O$_9^{2-}$ for the system. Due to the small Ir-Ir distance the direct hopping between the $a_1$ orbital (see SM) is appreciable (-573.9 meV) 
which is also reflected in the large $t_{2g}$ band width ($\sim1.88$ eV) of Ba$_3$YIr$_2$O$_9$.
\par
\begin{figure}[t]
\vspace{1.2cm}
\centering
\includegraphics[scale = .25]{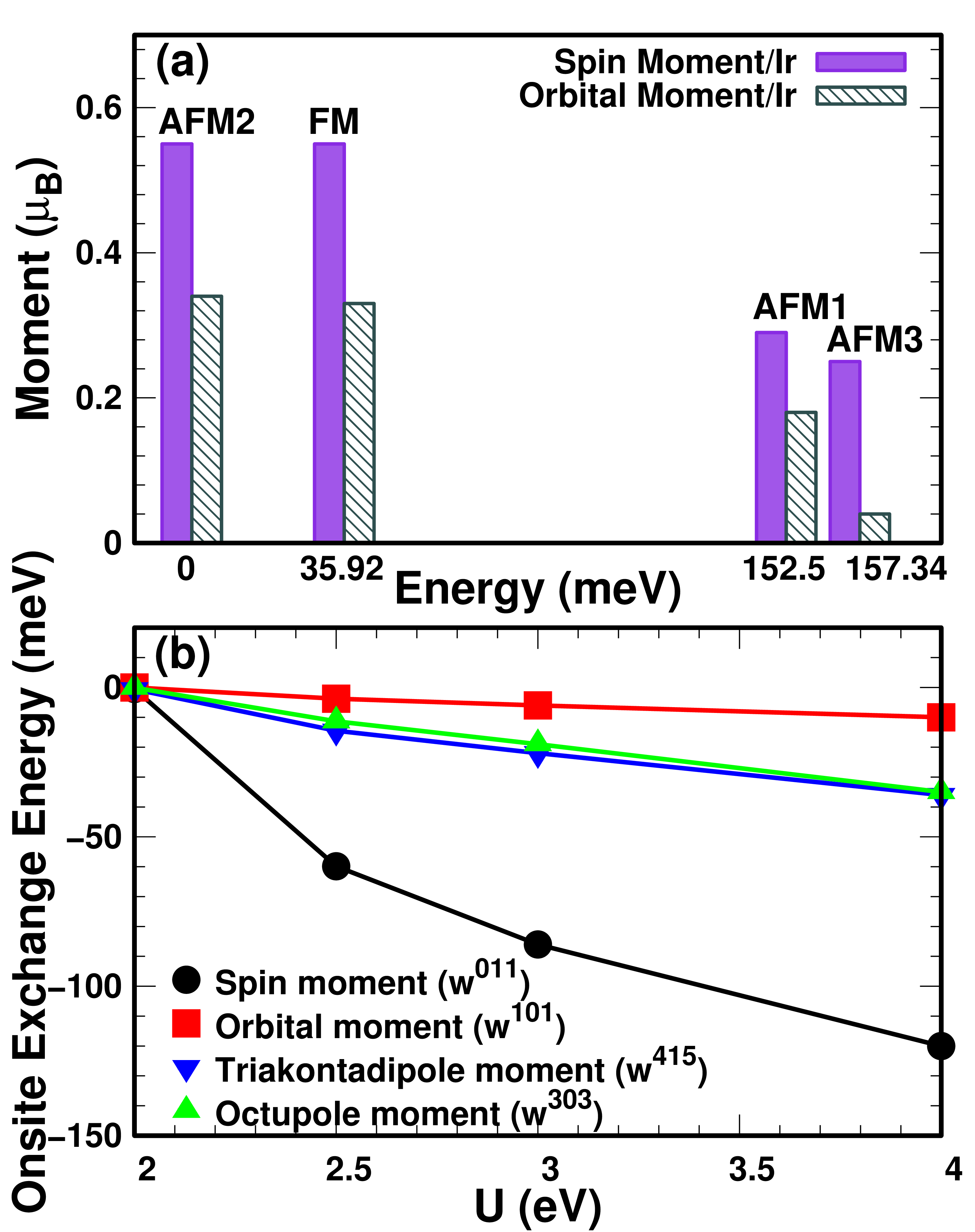}
\caption{(color online) (a) The variation of spin and orbital moment at the Ir site for various magnetic configurations in the AP phase. (b) Variation of onsite exchange energy associated with different order parameters (OP) with the variation of Hubbard U.}
\label{fig2}
\end{figure}

\begin{figure}[t]
\centering
\vspace{5.8 cm}
\includegraphics[scale =.25]{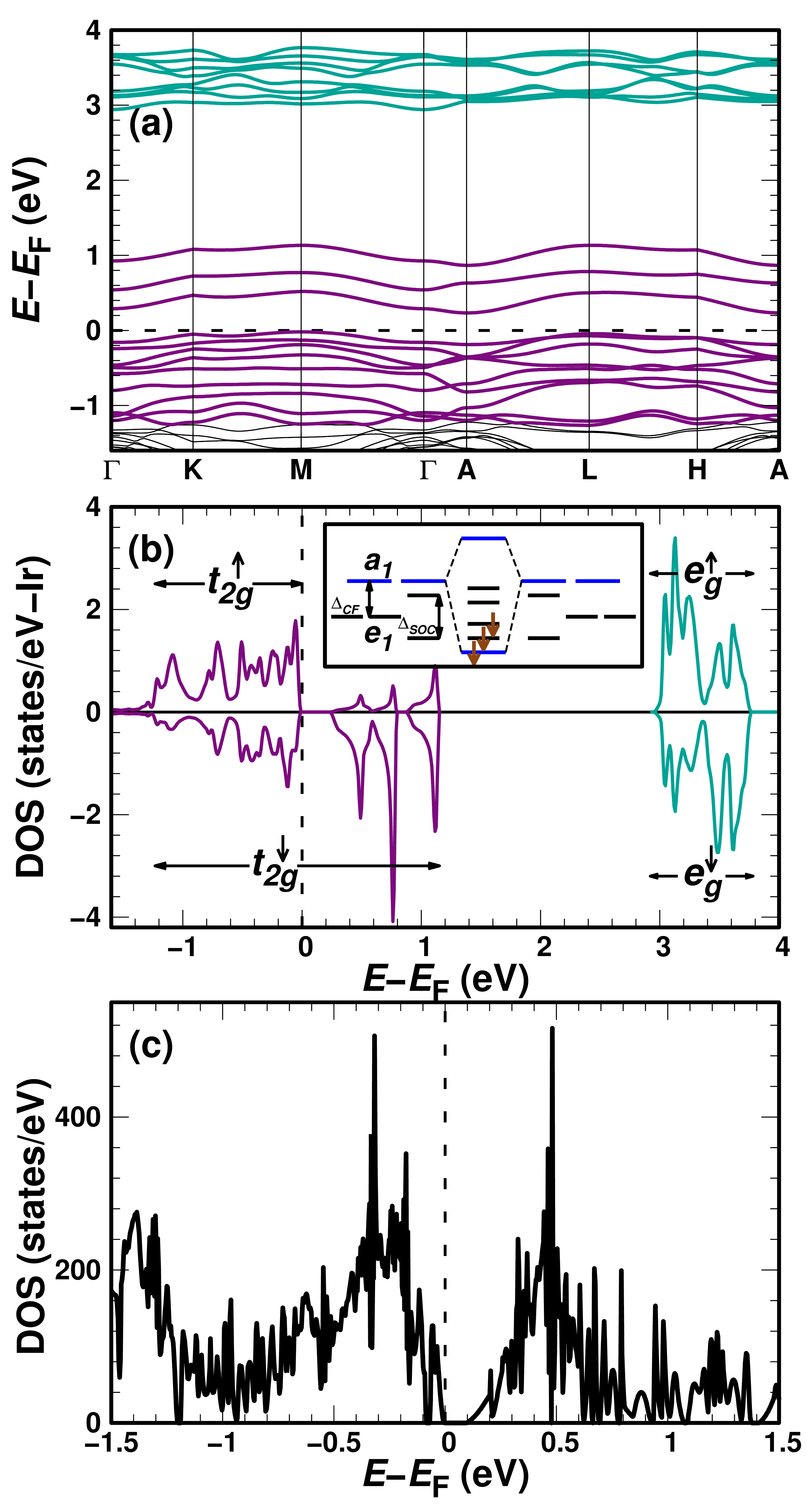}
\caption{(color online) (a) Bandstructure and (b) Partial density of states (DOS) for Ir-$d$ projected onto up and down spin state for the AP phase. The inset shows the schematic diagram to explain the insulating behavior of AP phase.  (c) The nonmagnetic DOS for the HP phase of Ba$_3$YIrO$_9$ in the presence of SOC and Hubbard $U$.}
\label{fig3}
\end{figure}

  
In order to understand the magnetic properties of the system four different magnetic configurations,
 namely FM (both the intra- and inter-dimer couplings are ferromagnetic), AFM1 (both the intra- and inter-dimer
couplings are antiferromagnetic), AFM2 (intra-dimer coupling is ferromagnetic and inter-dimer coupling is antiferromagnetic), and 
AFM3 (intra-dimer coupling is antiferromagnetic and inter-dimer coupling is ferromagnetic) have been simulated (See Fig. 2 of SM).   
The results of our LDA+U+SOC calculation with $U$ = 4 eV as obtained using the ELK code  are shown in Fig~\ref{fig2}(a).
We observe that the AFM2 state has the lowest energy with the spin (orbital) moment on the Ir site to be 0.55 (0.34) $\mu_B$. Spin and orbital moments are in the same direction as $t_{2g}$ states are more than half filled. The relatively smaller value of $\frac{m_l}{m_s}$ ratio ($\sim$0.62) compared to the other SOC driven $j_{eff}$ = 1/2 iridates, {\it i.e}, Sr$_2$IrO$_4$~\cite{Kim1}, (where it is $\sim$2) indicates the action of only moderately strong SOC in Ba$_3$YIr$_2$O$_9$. Fig.~\ref{fig2}(a) also reveal that the magnitude of Ir magnetic moments fluctuate for parallel and antiparallel spin configurations in a dimer and only the AFM2 configuration is insulating as illustrated by the plot of band structure and DOS shown in Fig.~\ref{fig3}(a) and ~\ref{fig3}(b) respectively. Although the LDA+U method is less accurate
than, the LDA method for total energy calculations~\cite{stoner_factor}
and has a tendency to favor magnetism because of the additional effective
Stoner coupling, it can still be used as a reference for magnetic
states as discussed in Ref.~\onlinecite{stoner_factor}.
 Therefore, a comparison of total energies among various magnetic configurations for Ba$_3$YIr$_2$O$_9$ should be reliable in a LDA+U+SOC calculation.
The stability of the ferromagnetic (FM) arrangement of Ir atoms in a dimer in the AFM2 configuration indicates a strong influence
of the effective Hund's coupling ($J_H$) in the system
 (which is directly related to
the onsite exchange parameter $J$), pointing to the fact that $J_H$
is not only larger than $\Delta_{CF}$ but also dominates over
 the splitting due to SOC ($\Delta_{SOC}$). This hierarchy of energies
 helps to promote Ir-Ir double exchange.
 This is in contrast to Ba$_5$AlIr$_2$O$_{11}$~\cite{Ba5AlIr2O11_arxiv},
 where SOC overcomes the Ir-Ir double exchange leading to an AFM arrangement of Ir spins in a dimer. 
\par  
The origin of the insulating state in Ba$_3$YIr$_2$O$_9$ can be explained by the cooperative effect of $J_{H}$, $\Delta_{CF}$, $\Delta_{SOC}$ and $t_{intra-dimer}$. In view of comparatively large $J_{H}$,  we ignore the spin mixing term of the SOC Hamiltonian~\cite{FeCr2S4} and only concentrate on the minority $t_{2g}$ bands. 
This simplification, allows us to diagonalise the SOC Hamiltonian, having matrix element $\langle i|\lambda \overline{L}\cdot \overline{S}|j\rangle$, (i,j~=~1,2,3) and find that the degeneracy of $|1\rangle$ and $|2\rangle$ (i.e, $e_1$ orbitals) is lifted upon inclusion of SOC, while $|3\rangle$ ($a_1$ orbital) remains unperturbed [see inset of Fig.~\ref{fig3}(b)]. In the presence of strong $t_{intra-dimer}$ (See Table-I of SM), the states will interact strongly with each other and form six non-degenerate states.
 Three electrons available for a pair of Ir in a dimer in the minority spin channel will lead to an insulating state. Further inclusion of finite Hubbard $U$ (4 eV), which is also crucial to stabilize magnetism, promotes an insulating state with an energy gap of 252 meV as shown in Fig.~\ref{fig3}(b).  
The calculated total magnetic moment in the AFM2 phase is substantially higher from that expected for a quarter filled $j_{eff}$=1/2 system, indicating $j_{eff}$=1/2 model may not be relevant here.\\


The above discussion places Ba$_3$YIr$_2$O$_9$ in the intermediate SOC regime and may be attributed to the large $t_{2g}$ band width that substantially renormalizes the strength of the atomic $\lambda$. In order to clarify this point further as well as to identify the order parameter (OP) responsible for breaking the time reversal (TR) symmetry that lead to magnetic order, we have carried out a multipolar (MP) analysis 
where the rotational invariant local Coulomb interaction is expanded in multipole tensors in the mean field limit. In the present case with Ir-d orbitals the density matrix D ($D=\left<d^{\phantom{\dagger}}_{n}d^{\dagger}_{n} \right>$, $d^{\dagger}_n$ and $d_{n}$ being creation and annihilation vector operators respectively for $d$-states of Ir at site $n$) has 100 independent elements. This information carried by the density matrix can be transformed to expectation values of multipole tensor moments through $w^{kpr}=\mathrm{Tr}\,\Gamma^{kpr}D $, where the multipolar tensor operator $\Gamma^{kpr}$ is an hermitian matrix-operator as given by Ref.~\onlinecite{Cricchio} and the indices are determined through the coupling of angular momenta, $0\le k\le 2\ell=4$, $0\le p \le 2s=1$ and $|k-p|\le r\le |k+p|$. Altogether there are 18 different multipole tensors (which in total has 100 independent tensor components), of which 9 multipole tensors break the TR symmetry.\\
In terms of these moments, $w^{kpr}$ the onsite exchange energy for the corresponding state ($E_x$) can be expanded as~\cite{Cricchio2}, 
\begin{align}
E_x=\frac{1}{2}\sum_{kpr} K_{kpr}w^{kpr} \cdot w^{kpr}
\end{align}
where, $K_{kpr}$ are the corresponding exchange parameters~\cite{Elk2}.
Thus the exchange energy (on site) associated with a multipole moment
$w^{kpr}$ can be identified~\cite{Cricchio3}. This corresponds to the stabilization of the concerned multipolar order.
\par
The onsite exchange energies corresponding to the TR odd multipole moments in the AFM2 state as a function of $U$ are shown in Fig.~\ref{fig2}(b) 
indicating no change in the hierarchy of the moments with $U$ (see also Table II of SM).
From Fig.~\ref{fig2}(b), we gather that the spin moment is the POP for the system as it is associated with the highest onsite exchange energy (-120 meV for U=4 eV).
This is in 
contrast to the case of SOC driven 
$j_{eff}$=1/2 Mott insulator Sr$_2$IrO$_4$~\cite{Sr2IrO4} with a higher order multipole (Triakontadipole) as POP where spin and orbital moments are entangled.

\par
Finally to check the possibility of a structural transition accompanied by a CO state, a complete structural relaxation calculation was performed (see SM). Calculations revealed environment of Ir-Ir dimer to be unaffected ruling out the possibility of a CO state very similar to Ba$_3$M$^{3+}$Ru$_2$O$_9$ systems~\cite{Ru_Y}.

 
\par 
Next we have analyzed the HP phase, considering a model structure (see SM), which has been described as a QSL state experimentally.
In view of the large size of the unit cell all calculations for the HP phase have been done in the plane wave basis as implemented in the VASP code~\cite{PAW1, PAW2, vasp1, vasp2}. 
In Fig~\ref{fig3}(c) we plot the total DOS for the paramagnetic phase in the
presence of SOC and U.  We find the system to be an insulator (for U$_{eff}$=U-J=1.5 eV) with an energy gap of 87 meV which is in agreement with the experiment~\cite{Spinliquid_cubicphase}.
 In order to have an estimate of the intersite magnetic interaction in the HP phase, we have considered three magnetic configurations, namely FM, and two antiferromagnetic configurations AFM-a and AFM-b 
 (see SM for description). The results of our calculation are shown in Table~\ref{table2}. In sharp contrast to the AP phase the energy difference between the magnetic configurations are small (see Table~\ref{table2}), 
 implying the small energy scale of the intersite exchange interaction ($J_{ex}$) in the HP phase. The AFM configurations being lower in energy, indicate the
 presence of a weak AFM interaction in the system, consistent with the small value of $\theta_{CW}$ (=-1.6 K) obtained experimentally~\cite{Spinliquid_cubicphase}. The weak $J_{ex}$ due to disorder prohibits long range order in the HP phase.

\begin{table}[t]
\caption{Energy difference between FM and two antiferromagnetic configurations 
along with the average moment ($\mu_B$/Ir). }
\centering\centering{}%
\begin{tabular}{|c|c|c|c|}
\hline
Configurations & $\Delta$E/Ir & Spin (Orb.) \\
               & (meV)        &  Moment/Ir\\
\hline
FM & 0.0 & 0.43 (0.14)\\
AFM-a & -0.41 & 0.40 (0.16)\\
AFM-b & -0.24 & 0.40 (0.16)\\
\hline
\end{tabular}
\label{table2}
\end{table}

\par
It is quite clear from the above discussion that there is a competition between SOC and $J_{ex}$ in the two 
phases of Ba$_3$YIr$_2$O$_9$. In the AP phase, while $J_{ex}$ 
wins over SOC so that $\frac{J_{ex}}{\lambda} > 1$,
the HP phase, on the other hand, has much weaker exchange interaction ($\sim$ few meV) compared to SOC (for
iridates this is typically $\sim$10$^2$ meV) making  $\frac{J_{ex}}{\lambda} < 1$. 
The latter condition in the HP phase of Ba$_3$YIr$_2$O$_9$ is very similar to the spinel compound FeSc$_2$S$_4$~\cite{FeSc2S4}.
Interestingly the model Hamiltonian considered for FeSc$_2$S$_4$ is also relevant for the cubic HP phase of Ba$_3$YIr$_2$O$_9$ as both of them have quite similar structure, therefore we expect in analogy, the stabilization of SOL state in the HP phase.
\par
In conclusion, our detailed study of electronic and magnetic properties of Ba$_3$YIr$_2$O$_9$ both in the AP and HP phase suggest that a competition between 
$J_{ex}$ and SOC decides the magnetic properties in both phases. In the AP phase the $J_{ex}$ dominates
over the SOC leading to long range magnetic order. The relatively weak value of SOC in the AP phase may be attributed to the large $t_{2g}$ band width in Ba$_3$YIr$_2$O$_9$.
The identification of the spin moment as the POP, also supports the fact that SOC is not so 
 strong in the AP phase. 
In the HP cubic phase of Ba$_3$YIr$_2$O$_9$ the SOC dominates over
the $J_{ex}$ prohibiting long range order. Finally, we propose the pressure induced modulation of $\frac{J_{ex}}{\lambda} < 1$ in the HP phase is responsible for the SOL state. 
\begin{acknowledgements}
The authors thank A. V. Mahajan, I. I. Mazin, and S. Streltsov for useful comments. I.D. thanks Department of Science and Technology (DST), Government of India for support. S.B. thanks Council of Scientific and Industrial Research (CSIR), India for a Fellowship. Y.L. and R.V. thank the Deutsche Forschungsgemeinschaft for financial support through grant TR/SFB 49. L.N. thanks Swedish Research Council (VR) for support.
\end{acknowledgements}


\newpage

\onecolumngrid{

\begin{center}
\large\bf{Supplementary Materials for \\
Electronic Structure and Spin-orbit driven Magnetism in $d^{4.5}$ insulator Ba$_3$YIr$_2$O$_9$}
\end{center}

\section{Computational Details}
Density functional theory calculations have been performed using four different methods, namely (a)
(APW + lo)method as implemented in ELK,
~\cite{Elk1,Elk2,Elk3}
(b)LAPW method as implemented in all electron full potential Wien2k code~\cite{wien2k},
(c) plane wave based method as implemented in Vienna \textit{ab initio} simulation package (VASP) ~\cite{vasp1,vasp2} and
(d) N$^{th}$ order muffin-tin orbital (NMTO) method~\cite{NMTO1, NMTO2, NMTO3}.
In order to investigate the details of the physics for the hexagonal ambient pressure (AP) phase of Ba$_3$YIr$_2$O$_9$ we have performed calculations with the augmented plane waves plus local orbitals (APW + lo) method within the local density approximation (LDA) as well as LDA+U+SOC including successively the effect of SOC and Hubbard $U$ as implemented in the code ELK.~\cite{Elk1,Elk2,Elk3} The U parameter was expressed as a linear combination of Slater parameters, which in turn were calculated as the radial integrals of the Yukawa screened Coulomb potential, and the localized limit was adopted for
the double counting correction.~\cite{Elk2} A 10$\times$10$\times$6 k-mesh has been used for the Brillouin-Zone integration.\\
The calculations were also performed with generalized gradient correction (GGA) of Perdew-Burke-Ernzerhof ~\cite{gga} including Hubbard U and SOC as implemented in the all electron full potential Wien2k~\cite{wien2k} code. We set the basis-size controlling parameter $RK_{max}$ equal to 8 and considered
a mesh of $9\times 9\times 3$ ${\bf k}$ points in the first Brillouin
zone (FBZ) for the self-consistency cycle.

We have also used a plane wave based method as implemented in Vienna \textit{ab-initio} simulation package (VASP)~\cite{vasp1,vasp2} where projector augmented wave potentials~\cite{PAW1,PAW2} are used to model the electron-ion interaction. A 10$\times$10$\times$6 k-mesh, similar to the ELK calculation, has been used for the Brillouin-Zone integration. Symmetry has been switched off in order to minimize possible numerical errors.
In order to check, whether there is any structural transformation of the AP phase to charge ordered state, upon relaxation we have carried out relaxation calulations within the framework of GGA+U+SOC method. All structural relaxations were carried out until the Hellman-Feynman forces became less than 0.01 eV/A$^\circ$. In this optimization, both the cell parameters and the positions of the atoms were allowed to relax. The experimentally synthesized HP cubic disordered structure is modelled by first constructing a 3$\times$3$\times$3 supercell of the ordered structure Ba$_2$YIrO$_6$ and then replacing 9 of the Y atoms by Ir atoms. In view of the large size of the unit cell the high pressure (HP) disordered cubic phase is studied only within GGA+U+SOC approach as implemented in VASP.\\
 Finally to extract the tight binding parameters and to understand the crystal field splitting, we have constructed low energy model Hamiltonian using the downfolding procedure as implemented in the NMTO method.~\cite{NMTO1, NMTO2, NMTO3} The NMTO method, for which the selfconsistent version is yet to be available, relies on the selfconsistent potentials borrowed from the linear MTO (LMTO)\cite{LMTO} calculations. For the self-consistent LMTO calculations within the atomic sphere approximation (ASA), no empty sphere was used for Ba$_3$YIr$_2$O$_9$. For the space filling MT radii used for Ba, Y, Ir, and O were
2.28 \AA, 1.50 \AA, 1.30 \AA, and 0.97 \AA , respectively.\\
\begin{figure}[h]
\centering
\includegraphics[scale=0.30]{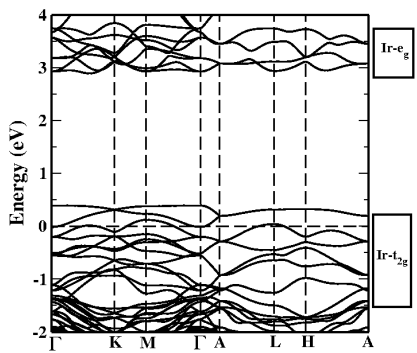}
\caption{LMTO Bandstructure of Ba$_3$YIr$_2$O$_9$ in absence of any magnetic order showing large t$_{2g}$-e$_g$ splitting.}
\label{Supp_fig1}
\end{figure}

\section{Crystal Field Splitting and estimation of the strength of the intra-dimer hopping for B\MakeLowercase{a}$_3$YI\MakeLowercase{r}$_2$O$_9$ }

In order to have a quantitative estimate of the intra dimer hopping strength and also to have an estimate of the various Ir $d$ levels,
we have carried out NMTO downfolding calculations for Ba$_3$YIr$_2$O$_9$ and constructed a Ir-d only
low-energy Hamiltonian by integrating out the high energy
degrees of freedom other than Ir-d. Diagonalization of the on-site block for Ba$_3$YIr$_2$O$_9$ gives the following five eigen states
with energies $-4.26$, $-4.26$,$-4.11$, $-0.48$ and $-0.48$ eV respectively,
\begin{eqnarray*}
|1\rangle & = & -0.8135|xy\rangle+0.5815|yz\rangle\\
|2\rangle & = & -0.5815|xz\rangle-0.8135|x^{2}-y^{2}\rangle\\
|3\rangle & = & |z^{2}\rangle\\
|4\rangle & = & -0.5815|xy\rangle-0.8135|yz\rangle\\
|5\rangle & = & -0.8135|xz\rangle+0.5815|x^{2}-y^{2}\rangle
\end{eqnarray*}

Clearly the first three states ($|1>$, $|2>$, and $|3>$) form the t$_{2g}$ manifold and the other two form the e$_{g}$ manifold. From the calculated energy eigenvalues it is clear that the t$_{2g}$ manifold splits into low lying doubly degenerate states $|1>$ and $|2>$ known as e$_1$ orbitals and singly degenerate state $|3>$ known as a$_1$ orbital. The t$_{2g}$-e$_{g}$ crystal field splitting is quite large ($\sim3.6$eV) for Ba$_3$YIr$_2$O$_9$ which is also clear from Fig~\ref{Supp_fig1}.
We have displayed the intra-dimer hopping between the Ir t$_{2g}$ states in Table~\ref{table:hopping} for Ba$_3$YIr$_2$O$_9$.

\begin{table}
\caption{The hopping parameters (in meV) between the t$_{2g}$ orbitals of the Ir atoms in the nonmagnetic phase of Ba$_3$YIr$_2$O$_9$. }
\centering
\begin{tabular}{|ll|l|l|l|l|l|}
\hline
Ir2 &  & $|1>$ & $|2>$ & $|3>$ \\[1 ex]
\hline
Ir1                 & $<1|$ & -213.9  & 0.0 & 0.0 \\
                    & $<2|$ & 0.0     & -213.9 & 0.0 \\
                    & $<3|$ & 0.0     & 0.0 & -573.9 \\
\hline
\end{tabular}
\label{table:hopping}
\end{table}

It is clear from Table~\ref{table:hopping}, that the direct hopping between the a$_1$ orbital is quite strong indicating the presence of strong intra dimer interaction in Ba$_3$YIr$_2$O$_9$.\\
\begin{figure}[h]
\centering
\includegraphics[scale=0.30]{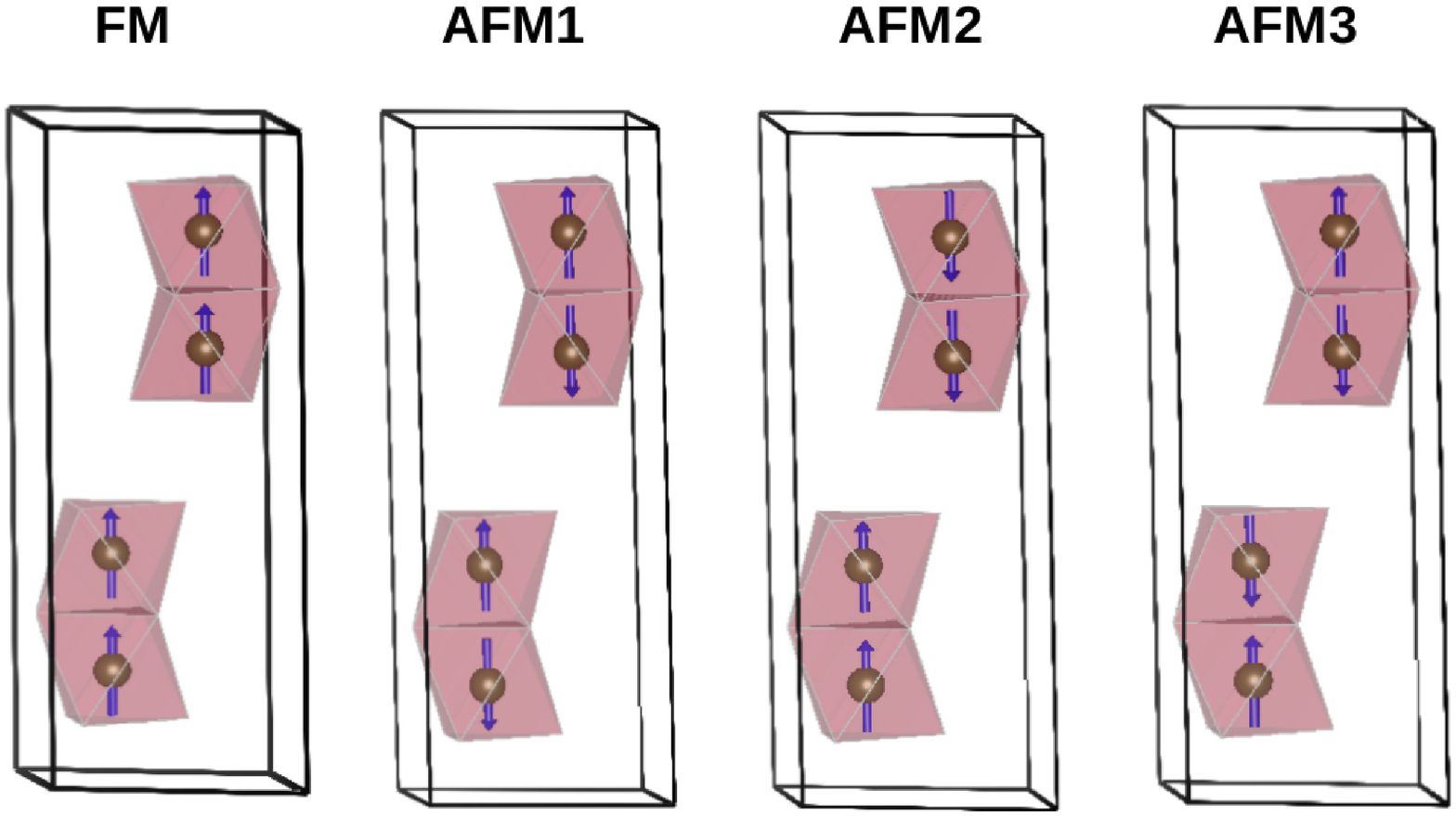}
\caption{Different magnetic configuration of Ba$_3$YIr$_2$O$_9$ in AP phase.}

\label{Supp_fig4}
\end{figure}

\begin{figure}[h]
\centering
\includegraphics[scale=0.60]{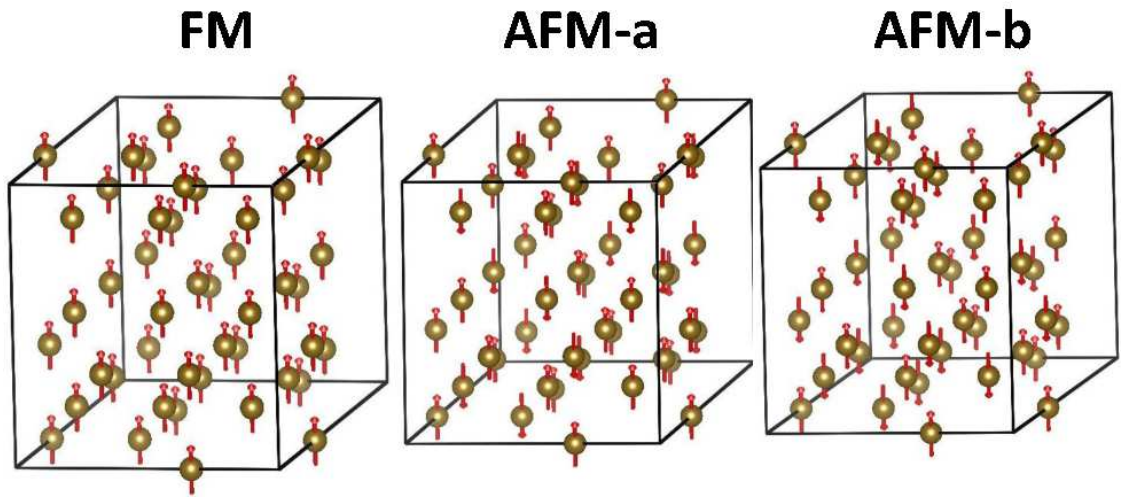}
\caption{Different magnetic configuration of Ba$_3$YIr$_2$O$_9$ in the HP phase}
\label{Supp_fig5}
\end{figure}

\section{Magnetic configurations}

The total energy calculations are carried out for the ferromagnetic (FM) and three antiferromagnetic configurations namely AFM1, AFM2 and AFM3 for the AP phase of Ba$_3$YIr$_2$O$_9$. Fig~\ref{Supp_fig4} shows the spin structure corresponding to FM, AFM1, AFM2 and AFM3. The dominant TR odd multipole moments and onsite exchange energies associated with the AFM2 state are shown in Table~\ref{tab2}.
\begin{table}
\centering
\caption{\label{tab2} Study of the variation of band gap and Order Parameters (OP) with $U$ for the AFM2 magnetic configuration.
 Significant magnetic OP like spin moment ($w^{011}$), orbital moment ($w^{101}$), triakontadipole moment ($w^{415}$) and octupole moment ($w^{303}$) are studied. }
\begin{tabular}{|c|c|c|c|c|c|c|}
\hline
  $\{U;J\}$ & Band Gap &  $w^{101}$  & $w^{011}$ & $w^{415}$ & $w^{303}$\\
  (in eV) & (in meV) & [Ex Eng] &  [Ex Eng]   &  [Ex Eng]     & [Ex Eng] \\
          &          & (in meV)  &  (in meV)    &  (in meV)  &  (in meV)  \\
\hline
 $\{4;1\}$ & 252.42 & 0.34 & 0.55 & 1.96 & 0.73 \\
           &  & [-10.0] & [-120] & [-36] & [-35] \\
 $\{3;0.94\}$ & 73.05 & 0.30 & 0.50 & 1.77 & 0.67 \\
             & &  [-6] & [-86] & [-22] & [-19] \\
 $\{2.5;0.88\}$  & 0.0 &0.26 & 0.45 & 1.55 & 0.58 \\
             & & [-3.78] & [-59.83] & [-14.49] & [-11.35] \\
$\{2;0.8\}$ & 0.0 & 0.02 & 0.04 & 0.35 & 0.06\\
             & & [0.0] & [-0.5] & [-0.6] & [-0.09] \\
\hline
\end{tabular}
\end{table}
For the HP phase of Ba$_3$YIr$_2$O$_9$ the total energy calculations are carried out for the magnetic configurations FM, AFM-a and AFM-b. Fig~\ref{Supp_fig5} shows the spin structure corresponding to these configurations. In the FM configuration all the spins on the 36 Ir atoms, present in the supercell, are aligned along (001)-axis, making all the Ir-Ir magnetic exchange interactions ferromagnetic.
However AFM-a and AFM-b configurations are made in such a way that some of the Ir spins are
pointed along (001)-axis while the rest are pointed along (00-1)-axis, which in turn make the presence of
both ferromagnetic and antiferromagnetic Ir-Ir interactions in these two configurations as illustrated in Fig.~\ref{Supp_fig5}. It is important to note that while the magnetic
configurations are markedly different the energy difference between the magnetic configurations are
small indicating a small energy scale of the intersite exchange interaction (see table-I in the main text).\\


\begin{thebibliography}{1}
 \bibitem{Sr2IrO4_Insulator} G. Cao, J. Bolivar, S. McCall, J. E. Crow, and R. P. Guertin, Phys. Rev. B {\bf 57}, R11039(R) (1998).

 \bibitem{Ba2IrO4_Insulator} R. Arita, J. Kuneˇs, A. V. Kozhevnikov, A. G. Eguiluz, and M. Imada, Phys. Rev. Lett. {\bf 108},
086403 (2012).

\bibitem{Na2IrO3} Y. Singh and P. Gegenwart, Phys. Rev. B {\bf 82}, 064412 (2010).

\bibitem{QSL_Na4Ir3O8} M. J. Lawler, A. Paramekanti, Y. B. Kim, and L. Balents, Phys. Rev. Lett. {\bf 101}, 197202 (2008).

\bibitem{QSL_Ba3IrTi2O9}T. Dey, A. V. Mahajan, P. Khuntia, M. Baenitz, B. Koteswararao, and F. C. Chou, Phys. Rev. B {\bf 86}, 140405(R) (2012).

\bibitem{TI_Na2IrO3_1} A. Shitade, H. Katsura, J. Kuneˇs, X.-L. Qi, S.-C. Zhang, and N. Nagaosa, Phys. Rev. Lett. {\bf 102}, 256403 (2009).

\bibitem{TI_Na2IrO3_2} H.-C. Jiang, Z.-C. Gu, X.-L. Qi, and S. Trebst, Phys. Rev. B {\bf 83}, 245104 (2011).
\bibitem{IrO2} S. K. Panda, S. Bhowal, A. Delin, O. Eriksson, and I. Dasgupta, Phys. Rev. B {\bf 89}, 155102 (2014).

\bibitem{Kim1} B. J. Kim, H. Jin, S. J.Moon, J.-Y. Kim, B.-G. Park, C. S. Leem, Jaejun Yu, T. W. Noh, C. Kim, S.-J. Oh, J.-H. Park, V. Durairaj, G. Cao, and E. Rotenberg, Phys. Rev. Lett. {\bf 101}, 076402 (2008).

\bibitem{Kim2} B. J. Kim, H. Ohsumi, T. Komesu, S. Sakai, T.Morita, H. Takagi, and T. Arima, Science {\bf 323}, 1329 (2009).


\bibitem{ZnIr2} A. Nag {\it et. al},  arXiv:1506.04312v3.

\bibitem{Spinliquid_cubicphase} Tusharkanti Dey, A. V. Mahajan, R. Kumar, B. Koteswararao, F. C. Chou, A. A. Omrani, and
H. M. Ronnow, Phys. Rev. B {\bf 88}, 134425 (2013).

\bibitem{RuCharge_ordering} Simon A. J. Kimber, Mark S. Senn, Simone Fratini, Hua Wu, Adrian H. Hill, Pascal Manuel, J. Paul Attfield, Dimitri N. Argyriou, and Paul. F. Henry, Phys. Rev. Lett. {\bf 108}, 217205 (2012).

\bibitem{Doi} Yoshihiro Doi and Yukio Hinatsu, J. Phys.: Condens. Matter {\bf 16} 2849 (2004).

\bibitem{TDey} Tusharkanti Dey, R. Kumar, A. V. Mahajan, S. D. Kaushik, and V. Siruguri, Phys. Rev. B {\bf 89}, 205101 (2014).

\bibitem{Elk1} D. Singh and L. Nordstr\"om, Planewaves, Pseudopotentials, and the LAPW method (Springer, New York, 2006).

\bibitem{Elk2} F. Bultmark, F. Cricchio, O. Gr\aa n\"{a}s, and L. Nordstr\"om, Phys. Rev. B {\bf 80}, 035121 (2009).

\bibitem{Elk3} ELK is available at http://elk.sourceforge.net.

\bibitem{gga}J. P. Perdew, K. Burke, and M. Ernzerhof, Phys. Rev. Lett. {\bf 77}, 3865 (1996).

\bibitem{wien2k} P. Blaha, K. Schwarz, G. K. H. Madsen, D. Kvasnicka,J. Luitz, WIEN2k, {\it An Augmented Plane Wave + Local Orbitals Program for Calculating Crystal Properties} (Karlheinz Schwarz, Techn. Universiat Wien, Austria, 2001.
\bibitem{PAW1} P. E. Bl\"{o}chl, Phys. Rev. B {\bf 50}, 17953 (1994).

\bibitem{PAW2} G. Kresse and D. Joubert, Phys. Rev. B {\bf 59}, 1758 (1999).

\bibitem{vasp1}G. Kresse and J. Hafner, Phys. Rev. B {\bf 47}, 558(R) (1993).

\bibitem{vasp2}G. Kresse and J. Furthm\"uller, Phys. Rev. B {\bf 54}, 11169 (1996).


\bibitem{NMTO1}O. K. Andersen and T. Saha-Dasgupta, Phys. Rev. B {\bf 62}, R16219(R) (2000).

\bibitem{NMTO2} O. K. Andersen, T. Saha-Dasgupta, R. W. Tank, C. Arcangeli, O. Jepsen, and G. Krier, Electronic Structure
and Physical Properties of Solids. The Uses of the LMTO Method, Springer Lecture Notes in Physics (Berlin:
Springer), 3 (2000).

\bibitem{NMTO3} O. K. Andersen, T. Saha-Dasgupta, and S. Ezhov, Third-generation muffin-tin orbitals, Bull. Mater. Sci. {\bf 26}, 19 (2003).

\bibitem{LMTO} O. K. Andersen and O. Jepsen, Phys. Rev. Lett. {\bf 53}, 2571 (1984).

\bibitem{SM} Supplementary Materials for theoretical techniques.

\bibitem{stoner_factor} A. G. Petukhov, I. I. Mazin, L. Chioncel and A. I. Lichtenstein, Phys. Rev. B {\bf 67}, 153106 (2003).

\bibitem{Ba5AlIr2O11_arxiv} S.V. Streltsov, J. Terzic, J. C. Wang, Feng Ye, D.I. Khomskii, W. H. Song, S. J. Yuan, S. Aswartham, and G. Cao, arXiv:1505.00877.

\bibitem{FeCr2S4} Soumyajit Sarkar, Molly De Raychaudhury, I. Dasgupta, and T. Saha-Dasgupta, Phys. Rev. B {\bf 80}, 201101(R) (2009).

\bibitem{Cricchio} F. Cricchio, O. Gr\aa n\"{a}s, and L. Nordstr\"{o}m, Eur. Phys. Lett. {\bf 94}, 57009 (2011).

\bibitem{Cricchio2} F. Cricchio, F. Bultmark, and L. Nordstr\"{o}m, Phys. Rev. B {\bf 78}, 100404(R) (2008).


\bibitem{Cricchio3} F. Cricchio, F. Bultmark, O. Gr{\aa}n\"as, and Lars Nordstr\"om, {\prl} {\bf 103}, 107202 (2009).

\bibitem{Sr2IrO4} Shreemoyee Ganguly, Oscar Gr\aa n\"{a}s, and Lars Nordstr\"om, Phys. Rev. B {\bf 91}, 020404(R) (2015).

\bibitem{Ru_Y} Mark S. Senn, Simon A. J. Kimber, Angel M. Arevalo Lopez, Adrian H. Hill, and J. Paul Attfield,
Phys. Rev. B {\bf 87}, 134402 (2013).

\bibitem{FeSc2S4} Gang Chen, Leon Balents, and Andreas P. Schnyder, Phys. Rev. Lett. {\bf 102}, 096406 (2009); Gang Chen, Andreas P. Schnyder and Leon Balents, Phys. Rev. B {\bf 80}, 224409 (2009); S. Sarkar, T. Maitra, Roser Valent\'i, and T. Saha-Dasgupta, Phys. Rev. B {\bf 82}, 041105(R) (2010).


\end{thebibliography}

\begin{thebibliography}{1}

\bibitem{Elk1} D. Singh and L. Nordstr\"om, Planewaves, Pseudopotentials, and the LAPW method (Springer, New York, 2006).

\bibitem{Elk2} F. Bultmark, F. Cricchio, O. Gr{\aa}n\"as, and Lars Nordstr\"om, Phys. Rev. B {\bf 80}, 035121 (2009).

\bibitem{Elk3} ELK is available at http://elk.sourceforge.net.

\bibitem{wien2k} P. Blaha, K. Schwarz, G. K. H. Madsen, D. Kvasnicka,J. Luitz, WIEN2k, {\it An Augmented Plane Wave + Local Orbitals Program for Calculating Crystal Properties} (Karlheinz Schwarz, Techn. Universiat Wien, Austria, 2001.

\bibitem{vasp1}G. Kresse and J. Hafner, Phys. Rev. B {\bf 47}, 558 (1993).

\bibitem{vasp2}G. Kresse and J. Furthmüller, Phys. Rev. B {\bf 54}, 11169 (1996).


\bibitem{NMTO1}O. K. Andersen and T. Saha-Dasgupta, Phys. Rev. B 62, R16219.
(2000).

\bibitem{NMTO2} O. K. Andersen, T. Saha-Dasgupta, R. W. Tank, C. Arcangeli, O. Jepsen, and G. Krier, Electronic Structure
and Physical Properties of Solids. The Uses of the LMTO Method, Springer Lecture Notes in Physics (Berlin:
Springer) , 3 (2000).

\bibitem{NMTO3} O. K. Andersen, T. Saha-Dasgupta, and S. Ezhov, Third-generation muffin-tin orbitals, Bull. Mater. Sci. 26, 19 (2003).

\bibitem{gga}J. P. Perdew, K. Burke, and M. Ernzerhof, Phys. Rev. Lett. {\bf 77}, 3865 (1996).

\bibitem{PAW1} P. E. Bl\"{o}chl, Phys. Rev. B {\bf 50}, 17953 (1994).

\bibitem{PAW2} G. Kresse and D. Joubert, Phys. Rev. B {\bf 59}, 1758 (1999).


\bibitem{LMTO} O. K. Andersen and O. Jepsen, Phys. Rev. Lett. {\bf 53}, 2571 (1984).

\end{thebibliography}
\end{document}